\documentclass[12pt]{article}
\usepackage{amsmath,color,authblk,empheq}
\input amssym
\input amssym.def

\def\red#1{{\color{red} #1}}
\begin{document}

\def\prg#1{\medskip\noindent{\bf #1}}  \def\ra{\rightarrow}
\def\lra{\leftrightarrow}              \def\Ra{\Rightarrow}
\def\nin{\noindent}                    \def\pd{\partial}
\def\dis{\displaystyle}                \def\inn{\hook}
\def\grl{{GR$_\Lambda$}}               \def\Lra{{\Leftrightarrow}}
\def\cs{{\scriptstyle\rm CS}}          \def\ads3{{\rm AdS$_3$}}
\def\Leff{\hbox{$\mit\L_{\hspace{.6pt}\rm eff}\,$}}
\def\bull{\raise.25ex\hbox{\vrule height.8ex width.8ex}}
\def\ric{{Ric}}                      \def\tric{{(\widetilde{Ric})}}
\def\tmgl{\hbox{TMG$_\Lambda$}}
\def\Lie{{\cal L}\hspace{-.7em}\raise.25ex\hbox{--}\hspace{.2em}}
\def\sS{\hspace{2pt}S\hspace{-0.83em}\diagup}   \def\hd{{^\star}}
\def\dis{\displaystyle}                 \def\ul#1{\underline{#1}}
\def\mb#1{\hbox{{\boldmath $#1$}}}     \def\grp{{GR$_\parallel$}}
\def\irr#1{^{(#1)}}                    \def\pha#1{\phantom{#1}}

\def\hook{\hbox{\vrule height0pt width4pt depth0.3pt
\vrule height7pt width0.3pt depth0.3pt
\vrule height0pt width2pt depth0pt}\hspace{0.8pt}}
\def\semidirect{\;{\rlap{$\supset$}\times}\;}
\def\first{\rm (1ST)}       \def\second{\hspace{-1cm}\rm (2ND)}
\def\bm#1{\hbox{{\boldmath $#1$}}}
\def\nb#1{\marginpar{{\large\bf #1}}}
\def\ir#1{{}^{(#1)}}  \def\ATan{\text{ArcTan}}

\def\G{\Gamma}        \def\S{\Sigma}        \def\L{{\mit\Lambda}}
\def\D{\Delta}        \def\Th{\Theta}
\def\a{\alpha}        \def\b{\beta}         \def\g{\gamma}
\def\d{\delta}        \def\m{\mu}           \def\n{\nu}
\def\th{\theta}       \def\k{\kappa}        \def\l{\lambda}
\def\vphi{\varphi}    \def\ve{\varepsilon}  \def\p{\pi}
\def\r{\rho}          \def\Om{\Omega}       \def\om{\omega}
\def\s{\sigma}        \def\t{\tau}          \def\eps{\epsilon}
\def\nab{\nabla}      \def\btz{{\rm BTZ}}   \def\heps{\hat\eps}
\def\bt{{\bar t}}     \def\br{{\bar r}}    \def\bth{{\bar\theta}}
\def\bvphi{{\bar\vphi}}   \def\ub#1{\underbrace{#1}}
\def\bx{{\bar x}}     \def\by{{\bar y}}     \def\bom{{\bar\om}}
\def\tphi{{\tilde\vphi}}  \def\tt{{\tilde t}} \def\bd{{\bar\d}}

\def\tG{{\tilde G}}   \def\cF{{\cal F}}      \def\bH{{\bar H}}
\def\cL{{\cal L}}     \def\cM{{\cal M }}     \def\cE{{\cal E}}
\def\cH{{\cal H}}     \def\hcH{\hat{\cH}}    \def\rd{\hat{\delta}}
\def\cK{{\cal K}}     \def\hcK{\hat{\cK}}    \def\cA{{\cal A}}
\def\cO{{\cal O}}     \def\hcO{\hat{\cal O}} \def\cV{{\cal V}}
\def\tom{{\tilde\omega}} \def\cS{{\cal S}}   \def\cE{{\cal E}}
\def\cR{{\cal R}}    \def\hR{{\hat R}{}}     \def\hL{{\hat\L}}
\def\tb{{\tilde b}}  \def\tA{{\tilde A}}     \def\tv{{\tilde v}}
\def\tT{{\tilde T}}  \def\tR{{\tilde R}}     \def\tcL{{\tilde\cL}}
\def\hy{{\hat y}\hspace{1pt}}  \def\tcO{{\tilde\cO}}
\def\bA{{\bar A}}     \def\bB{{\bar B}}      \def\bC{{\bar C}}
\def\bG{{\bar G}}     \def\bD{{\bar D}}      \def\bH{{\bar H}}
\def\bK{{\bar K}}     \def\bL{{\bar L}}      \def\rp{\hbox{(r1$^\prime$)}}

\def\rdc#1{\hfill\hbox{{\small\texttt{reduce: #1}}}}
\def\chm{\checkmark}  \def\chmr{\red{\chm}}
\def\nn{\nonumber}                    \def\vsm{\vspace{-9pt}}
\def\be{\begin{equation}}             \def\ee{\end{equation}}
\def\ba{\begin{align}}                \def\ea{\end{align}}
\def\bea{\begin{eqnarray} }           \def\eea{\end{eqnarray} }
\def\beann{\begin{eqnarray*} }        \def\eeann{\end{eqnarray*} }
\def\beal{\begin{eqalign}}            \def\eeal{\end{eqalign}}
\def\lab#1{\label{eq:#1}}             \def\eq#1{(\ref{eq:#1})}
\def\bsubeq{\begin{subequations}}     \def\esubeq{\end{subequations}}
\def\bitem{\begin{itemize}}           \def\eitem{\end{itemize}}
\renewcommand{\theequation}{\thesection.\arabic{equation}}
\title{Entropy in Poincar\'e gauge theory: Kerr-AdS solution}

\author{M. Blagojevi\'c and B. Cvetkovi\'c\footnote{
        Email addresses: \texttt{mb@ipb.ac.rs, cbranislav@ipb.ac.rs}}}
\affil{\normalsize{Institute of Physics, University of Belgrade,
                      Pregrevica 118, 11080 Belgrade, Serbia}}
\date{\today}
\maketitle
\begin{abstract}
Using a Hamiltonian approach, we introduce black hole entropy for Kerr-AdS spacetimes with torsion as the canonical charge on horizon. In spite of a completely different geometric setting with respect to GR, the resulting thermodynamic variables, energy, angular momentum and entropy, are shown to be proportional to the corresponding GR expressions. The validity of the first law is confirmed.
\end{abstract}

\section{Introduction}
\setcounter{equation}{0}

The entropy of black holes plays a crucial role in black hole thermodynamics. In the approach of Wald \cite{wald}, entropy is introduced in the framework of (Riemannian) diffeomorphism invariant theories as the Noether charge on horizon. In these theories, the gravitational dynamics is described by a metric of spacetme only, as is the case in general relativity (GR), and matter fields are tensor fields on spacetime manifold. After some time, Jacobson and Mohd \cite{JM} extended these considerations to theories whose spacetime geometry is described by an orthonormal coframe and the related Lorentz (or spin) connection. Staying close to the spirit of GR, they restricted their analysis to a torsionless Lorentz connection, which is completely determined in terms of the coframe field. Thus, in spite of the change of basic dynamical variables, the geometry of spacetime remained Riemannian.

A quite natural extension of the treatment of entropy was proposed recently in Ref. \cite{bc1}, where the Lorentz connection  was liberated from its Riemannian constraints by going over to Poincar\'e gauge theory (PG), a modern gauge-field-theoretic approach to gravity \cite{bh,pg}. In analogy to gauge theories of internal symmetries, PG is constructed by localizing the Poincar\'e group (translations and Lorentz rotations) of spacetime symmetries. In PG, the basic gravitational variables are again the coframe and the Lorentz connection, but here, in contrast to GR, the spacetime geometry is characterized by two types of field strengths, the torsion and the curvature.

The Hamiltonian approach to entropy proposed in Ref. \cite{bc1} describes the asymptotic charges (energy and angular momentum) and entropy as the canonical charges at infinity and horizon, respectively. It was successfully applied to spherically symmetric and asymptotically flat Kerr solutions in PG, and to the Kerr-Anti-de Sitter (Kerr-AdS) black holes in GR \cite{bc1,bc2,bc3}. Once the asymptotic charges and entropy are calculated, they are also shown to satisfy the first law of black hole thermodynamics, which is an independent test of the formalism. The objective of the present paper is to extend our Hamiltonian approach to physically more interesting but technically rather involved case of the Kerr-AdS black hole with torsion \cite{kads1,kads2}, see also \cite{kads3,kads4}.

The paper is organized as follows. In section  \ref{sec2}, we describe basic aspects of our Hamiltonian approach to entropy in PG, and section \ref{sec3} offers a review of the geometry of Kerr-AdS spacetimes with torsion. Then, in section \ref{sec4}, we apply the Hamiltonian approach to calculate energy and angular momentum of the Kerr-AdS black hole in PG, with  respect to the AdS background configuration. In this analysis, a particular attention is payed to a proper treatment of the Boyer-Lindquist coordinate system in the asymptotic region. Section \ref{sec5} is the central part of the present paper as it contains a detailed derivation of the Kerr-AdS black hole entropy. In section \ref{sec6}, we give a short verification of the validity of the first law of black hole thermodynamics, and section \ref{sec7} is devoted to concluding remarks. Finally, three appendices contain some technical details of our analysis of entropy.

Our conventions are the same as in Refs. \cite{bc1,bc2,bc3}. The Latin indices $(i,j,\dots)$ are the local Lorentz indices, the Greek indices $(\m,\n,\dots)$ are the coordinate indices, and both run over $0,1,2,3$. The orthonormal coframe (tetrad) is $b^i=b^i{}_\m dx^\m$, the dual basis (frame) is $h_i=h_i{}^\m\pd_\m$, $\om^{ij}=\om^{ij}{}_\m dx^\m$ is the Lorentz connection, the metric components in the local Lorentz and coordinate basis are $\eta_{ij}=(1,-1,-1,-1)$ and $g_{\m\n}=g_{ij} b^i{}_\m b^j{}_\n$, respectively, and $\ve_{ijmn}$ is the totally antisymmetric symbol with $\ve_{0123}=1$. The Hodge dual of a form $\alpha$ is denoted by $\hd\alpha$, and the wedge product of forms is implicitly understood.

\section{Entropy as the canonical charge}\label{sec2}
\setcounter{equation}{0}

To prepare our analysis of entropy for Kerr-AdS black holes with torsion, we start with a short account of the (geometric and) dynamical structure of PG \cite{bh,pg} and the basic aspects of the Hamiltonian understanding of black hole entropy \cite{bc1,bc2,bc3}.

The geometric structure of spacetime in PG is characterised by the existence of two gauge potentials, the coframe (tetrad) $b^i$ and the Lorentz connection $\om^{ij}=-\om^{ji}$ (1-forms), the related field strengths are the torsion $T^i:=d b^i+\om^i{}_k b^k$ and the curvature $R^{ij}:=d\om^{ij}+\om^i{}_k\om^{kj}$ (2-forms), and the associated spacetime structure is described by a Riemann-Cartan (RC) geometry.

The PG dynamics is determined by a Lagrangian $L=L_G+L_M$ (4-form), where $L_G$ is the pure gravitational part and $L_M$ describes matter fields and their gravitational interactions. The gravitational Lagrangian is assumed to be parity invariant and at most quadratic in the field strengths:
\be
L_G=-\hd(a_0R+2\L)+T^i\sum_{n=1}^3\hd(a_n\ir{n}T_i)
             +\frac{1}{2}R^{ij}\sum_{n=1}^6\hd(b_n\ir{n}R_{ij})\,,   \lab{2.1}
\ee
where $(a_0,\L,a_n,b_n)$ are the coupling constants, and $\ir{n}T_i,\ir{n}R_{ij}$ are irreducible parts of the field strengths, see, for instance, Ref. \cite{bc1}.
The variation of $L_G$ with respect to $b^i$ and $\om^{ij}$ yields the gravitational field equations in vacuum. After introducing the covariant gravitational momenta $H_i:=\pd L_G/\pd T^i$ and $H_{ij}:=\pd L_G/\pd R^{ij}$ (2-forms), and the associated energy-momentum and spin currents,
$E_i:=\pd L_G/\pd b^i$ and $E_{ij}:=\pd L_G/\pd\om^{ij}$ (3-forms), the gravitational field equations take a compact form
\bsubeq\lab{2.2}
\bea
\d b^i:&&\quad \nab H_i+E_i=0\, ,                                 \lab{2.2a}\\
\d\om^{ij}:&&\quad \nab H_{ij}+E_{ij}=0\,.                        \lab{2.2b}
\eea
\esubeq
In the presence of matter, the right-hand sides of \eq{2.2a} and \eq{2.2b} contain the corresponding matter currents.

The explicit expressions for the covariant momenta
\bsubeq\lab{2.3}
\bea
&&H_i=2\sum_{m=1}^2\hd(a_n \ir{m}T_i)\,,          \\
&&H_{ij}=-2a_0\hd(b_ib_j)+2\sum_{n=1}^6\hd(b_n\ir{n}R_{ij})\,,
\eea
\esubeq
play and important role in the analysis of black hole entropy.

The asymptotic conserved charges (energy and angular momentum) in PG are closely related to the regularity (functional differentiability) of the canonical gauge generator of local Poincar\'e symmetries. Following the ideas of Regge and Teitelboim \cite{rt1974}, the canonical form of these charges can be expressed in terms of certain surface integrals at spatial infinity, see Refs. \cite{kads3,bv1983,nester}. On the other hand, the concept of black hole entropy in GR is best understood as the \emph{Noether charge} on horizon \cite{wald}. As shown in Ref. \cite{bc1}, this idea can be quite naturally extended to PG by introducing entropy as the \emph{canonical charge} on horizon. By construction, this extension can be applied not only to black holes with torsion, but also to Riemannian black holes.

For a stationary black hole spacetime, its spatial section $\S$ is assumed to  have two components, one at infinity and the other at horizon, $\pd\S=S_\infty\cup S_H$. The corresponding boundary integral $\G$ has two parts, $\G=\G_\infty-\G_H$, which are determined by the following variational equations:
\bsubeq\lab{2.4}
\bea
&&\d\G_\infty=\oint_{S_\infty}\d B(\xi)\,,\qquad
       \d\G_H=\oint_{S_H} \d B(\xi)\,,                                  \\
&&\d B(\xi):=(\xi\inn b^{i})\d H_i+\d b^i(\xi\inn H_i)
   +\frac{1}{2}(\xi\inn\om^{ij})\d H_{ij}
   +\frac{1}{2}\d\om^{ij}(\xi\inn\d H_{ij})\, .
\eea
\esubeq
Here, $\xi$ is the Killing vector which takes the values  $\pd_t$ and/or $\pd_\vphi$ on $S_\infty$,  and becomes a linear combination thereof on $S_H$, such that $\xi^2=0$. The variation $\d B$ is determined in accordance with the \emph{boundary conditions},  which must be chosen so as to ensure the solutions for $\G_\infty$ and $\d\G_H$ to exist and be finite. In particular, $\d$ is required to satisfy the following rules:
\bitem
\item[(r1)] On $S_\infty$, the variation $\d$ acts on the parameters of a black hole solution, but not on the para\-me\-ters of the background configuration.\vsm
\item[(r2)] On $S_H$, the variation $\d$ must keep surface gravity constant.
\eitem
When the variational equations \eq{2.4} are \emph{$\d$-integrable} and the solutions for $\G_\infty$ and $\G_H$ are \emph{finite}, they are interpreted as the asymptotic charges and black hole entropy, respectively.

Although $\G_\infty$ and $\G_H$ are defined as a priory independent quantities,
the analysis of their construction \cite{bc1} reveals that the regularity of the canonical gauge generator is ensured by the relation
\be
\d\G\equiv\d\G_\infty-\d\G_H=0\,,                                   \lab{2.5}
\ee
which is equivalent to the first law of black hole thermodynamics.

\section{Kerr-AdS black hole with torsion}\label{sec3}
\setcounter{equation}{0}

In this section, we present Baekler et. al. Kerr-AdS solution \cite{kads1,kads2} in the framework of a wider class of parity even PG Lagrangians \cite{kads3}; for an extension to the general parity violating Lagrangian, see Obukhov \cite{kads4}.

\subsection{Metric and tetrad}

The metric of Kerr-AdS spacetime in Boyer-Lindquist coordinates takes the form \cite{carter,hentei,gibbons}
\bsubeq\lab{3.1}
\be
ds^2=\frac{\D}{\r^2}\Big(dt+\frac{a}{\a}\sin^2\th d\vphi\Big)^2
     -\frac{\r^2}{\D}dr^2-\frac{\r^2}{f}d\th^2
 -\frac{f}{\r^2}\sin^2\th\Big[a dt+\frac{(r^2+a^2)}{\a}d\vphi\Big]^2\,,\lab{3.1a}
\ee
where
\bea
&&\D(r):=(r^2+a^2)(1+\l r^2)-2mr\, ,\qquad \a:=1-\l a^2\,,              \nn\\
&&\r^2(r,\th):=r^2+a^2\cos^2\th\,,\qquad
  f(\th):=1-\l a^2\cos^2\th\,.
\eea
\esubeq
Here, $m$ and $a$ are the parameters of the solution, $\l=-\L/3a_0$, $\a$ normalizes the range of the angular variable $\vphi$ to $2\pi$, and $0\le\th<\pi$. For $m=0$, the metric reduces to the AdS form, albeit in somewhat ``twisted" coordinates \cite{carter,hentei}. The metric possesses two Killing vectors, $\pd_t$ and $\pd_\vphi$, and the larger root of $\D(r)=0$ defines the outher horizon,
\be
(r_+^2+a^2)(1+\l r_+^2)-2mr_+=0\,.
\ee
The angular velocity is given by
\be
\om(r):=\frac{g_{t\vphi}}{g_{\vphi\vphi}}
    =\frac{a\a\big[f(r^2+a^2)-\D\big]}{f(r^2+a^2)^2-a^2\D\sin^2\th}\, ,\qquad
\om(r_+)=\frac{a\a}{r_+^2+a^2}\, ,\chmr                              \lab{3.3}
\ee
Note that $\om(r)$ does not vanish for large $r$, $\om\sim -\l a+O_2$.
Surface gravity has the form
\be
\k=\frac{[\pd\D]_{r_+}}{2(r_+^2+a^2)}
  =\frac{r_+(1+\l a^2+3\l r_+^2-a^2/r_+^2)}{2(r_+^2+a^2)}\,.
\ee

The orthonormal tetrad associated to the metric \eq{3.1} is chosen in the form
\bea
&&b^0=N\Big(dt+\frac{a}{\a}\sin^2\th\,d\vphi\Big)\,,\qquad
  b^1=\frac{dr}{N}\,,                                                \nn\\
&&b^2=Pd\th\, ,\qquad
  b^3=\frac{\sin\th}{P}\Big[a\,dt+\frac{(r^2+a^2)}{\a}d\vphi\Big]\,, \lab{3.5}
\eea
where
\be
N(r,\th)=\sqrt{\D/\r^2}\, ,\qquad P(r,\th)=\sqrt{\r^2/f}\, .          \nn
\ee
A simple calculation of the horizon area yields
\be
A=\int_{r_+}b^2b^3=\frac{4\pi(r_+^2+a^2)}{\a}\,.
\ee
The Riemannian connection $\tom^{ij}$ is defined in the usual way as
\be
\tom^{ij}:=\frac{1}{2}\Big[h^i\inn d b^j-h^j\inn d b^i
                        -\Big(h^i\inn (h^j\inn d b^m)\Big)b_m\Big]\,,\lab{3.7}
\ee
see also appendix \ref{appA}.

\subsection{Torsion, connection and curvature}

The ansatz for torsion is given by \cite{kads2,kads3}
\bea
&&T^0=T^1=\frac{1}{N}\Big[-V_1b^0b^1-2V_4b^2b^3\Big]
          +\frac{1}{N^2}\Big[V_2b^-b^2+V_3b^-b^3\Big]\,,        \nn\\
&&T^2:=\frac{1}{N}\Big[ V_5b^-b^2+V_4b^-b^3\Big]\,,             \nn\\
&&T^3:=\frac{1}{N}\Big[-V_4b^-b^2+V_5b^-b^3\Big]\,,        \lab{3.8}
\eea
where $b^-:=b^0-b^1$ and the torsion functions $V_n$ have the form
\bea
&&V_1=\frac{m}{\r^4}(r^2-a^2\cos^2\th)\, ,\qquad
  V_2=-\frac{m}{\r^4 P}ra^2\sin\th\cos\th\, ,                       \nn\\
&&V_3=\frac{m}{\r^4 P}r^2a\sin\th\, ,\qquad
  V_4=\frac{m}{\r^4}ra\cos\th\,,\qquad V_5=\frac{m}{\r^4}r^2\,.
\eea
Thus, the torsion tends to zero at spatial infinity.
The irreducible components of $T^i$ are displayed in Appendix \ref{appA}; in particular, $\ir{3}T^i=0$. After introducing the contorsion 1-form,
\bsubeq
\be
K^{ij}:=\frac{1}{2}\Big[h^i\inn T^j-h^j\inn T^i
                  -\Big(h^i\inn\big( h^j\inn T^k\big)\Big) b_k\Big]\,,
\ee
or more explicitly
\bea
&&K^{01}=\frac{1}{N}V_1b^-\, ,                                    \nn\\
&&K^{02}=K^{12}=-\frac{1}{N^2}V_2b^-
                +\frac{1}{N}\big(V_5 b^2-V_4 b^3\big)\,,             \nn\\
&&K^{03}=K^{13}=-\frac{1}{N^2}V_3b^-
                +\frac{1}{N}\big(V_4 b^2+V_5 b^3\big)\, ,            \nn\\
&&K^{23}=-\frac{2}{N}V_4b^-\, ,
\eea
\esubeq
the RC connection is given by
\be
\om^{ij}=\tom^{ij}+K^{ij}\, .
\ee
\bitem
\item The tetrad field $b^i$ and the Lorentz connection $\om^{ij}$ are basic elements of the RC geometry of spacetime.
\eitem

The RC curvature $R^{ij}=d\om^{ij}+\om^i{}_k\om^{kj}$ has only two nonvanishing irreducible parts, $\ir{4}R^{ij}$ and $\ir{6}R^{ij}$;  with $A=(0,1)$ and $c=(2,3)$, they are given by
\be
\ir{6}R^{ij}=\l b^ib^j\, ,\qquad
             \ir{4}R^{Ac}=\frac{\l m r}{\D}b^-b^c\,.
\ee
The quadratic invariants
\be
R^{ij}\hd R_{ij}=12\l^2\heps\, ,\qquad T^i\hd T_i=0\,,
\ee
where $\heps:=b^0b^1b^2b^3$ is the volume 4-form, are regular. Note that the curvature invariant differs from its Riemannian analogue \cite{bc3}.

The effective form of the Lagrangian is determined by the nonvanishing irreducible parts of the field strengths,
\be
L_G=-\hd(a_0R+2\L_0)+T^i\hd(a_1\ir{1}T_i+a_2\ir{2}T_i)
           +\frac{1}{2}R^{ij}\hd(b_4\ir{4}R_{ij}+b_6\ir{6}R_{ij})\,.\lab{3.13}
\ee
The Kerr-AdS geometry is a solution of the PG field equations \eq{2.2} provided the Lagrangian parameters satisfy the following restrictions:
\be
2a_1+a_2=0\,,\qquad a_0-a_1-\l(b_4+b_6)=0\,,\qquad 3\l a_0+\L=0\, .
\ee
With the above form of $L_G$, the covariant momenta \eq{2.3} are determined by
\bea
&&H_i=2a_1\,\hd(\ir{1}T_i-2\,\ir{2}T_i)\, ,                          \nn\\
&&H_{ij}=-2(a_0-\l b_6)\,\hd(b_ib_j)+2b_4\hd\ir{4}R_{ij}\, ,
\eea
see also appendix \ref{appA}.

\section{Asymptotic charges}\label{sec4}
\setcounter{equation}{0}

As shown by Carter \cite{carter} and Henneaux and Teitelboim \cite{hentei}, Boyer-Lindquist coordinates are not adequate for analyzing the asymptotic charges of Kerr-AdS spacetime since the corresponding asymptotic behavior of the metric components is twisted with respect to the standard AdS backgrond configuration.  However, as we discussed in \cite{bc3}, one can use Boyer-Lindquist coordinates as a technically simple first step in the calculations, whereupon the transition to the new, ``untwisted" coordinates
\bea
T=t\, ,\qquad \phi=\vphi-\l at                                     \lab{4.1}
\eea
yields the correct final result. In fact, Henneaux and Teitelboim's analysis, based on the properties of asymptotic states, yields formulas for the new coordinates which also include an additional part transforming $(r,\th)$ into $(R,\Th)$. However, that part is not needed in our approach which is based on the Hamiltonian variational approach \eq{2.4}.

Under the coordinate transformation \eq{4.1}, the components od the Killing vector $\xi$  and the metric tensor $g_{\m\n}$ transform according to
\bea
&&\xi_T=\xi_t+\l a\xi_\vphi\, ,\qquad \xi_\phi=\xi_\vphi\,,          \nn\\
&&g_{T\phi}=g_{tt}+\l a g_{\vphi\vphi}\, ,\qquad
                              g_{\phi\phi}=g_{\vphi\vphi}\, ,        \nn\\
&&g_{TT}=g_{tt}+2\l a g_{t\vphi}+(\l a)^2g_{\vphi\vphi}\,.         \lab{4.2}
\eea

Before we begin with calculations, let us note that the background configuration, which is defined by $m=0$, also depends on the parameter $a$. Hence, in order to avoid the variation of those $a$'s that ``belong" to the background,  we introduce an improved interpretation of the rule (r1) formulated in section \ref{sec3}:
\bitem
\item[\rp] In the variational equation \eq{2.4} for $\d\G_\infty(\xi)$, first apply $\d$ to all the parameters $(m, a)$ appearing in $B(\xi)$, then subtract those $\d a$ terms that survive the limit $m = 0$, as they originate from the variation of the AdS background.
\eitem

In the calculations that follow, we use the notation
\bea
&&A_0:=a_0-\l(b_4+b_6)\equiv a_1\,,                                 \nn\\
&&d\Om:=\sin\th d\th d\vphi\to 4\pi\, ,\qquad
  d\Om':=\sin^3\th d\th d\vphi\to\frac{2}{3}4\pi\, ,
\eea
Various components of $\om^{ij}$ and $H_i,H_{ij}$ can be found with the help of Appendix A.

\subsection{Angular momentum}

We start the analysis of angular momentum by calculating the expression $\d E_\vphi:=\d\G_\infty(\pd_\vphi)$. For simplicity, we write $\d E_\vphi$ in the form $\d E_\vphi=\d E_{\vphi 1}+\d E_{\vphi 2}$, where
\bea
&&\d E_{\vphi 1}:=\frac{1}{2}\om^{ij}{}_\vphi\d H_{ij}
                    +\frac{1}{2}\d\om^{ij}H_{ij\vphi}\, ,         \nn\\
&&\d E_{\vphi 2}:=b^i{}_\vphi\d H_{i}+\d b^iH_{i\vphi}\,,
\eea
and the integration over $S_\infty$ is implicitly understood. The calculation is performed by ignoring $\d a$ terms that are independent of $m$, even when they are divergent, and by omitting asymptotically vanishing $O(r^{-n})$ terms. The nonvanishing contributions are given by
\bsubeq
\bea
\d E_{\vphi 1}&=&\om^{13}{}_\vphi\d H_{13}+\d\om^{13}H_{13\vphi}
   =\big(\om^{13}{}_\vphi\d H_{13\th\vphi}
        +\d\om^{13}{}_\vphi H_{13\th\vphi}\big)d\th d\vphi         \nn\\
  &=&\d\big(\om^{13}{}_\vphi H_{13\th\vphi}\big)d\th d\vphi
        =2A_0\d\Big(\frac{ma}{\a^2}\Big)d\Om'\,,                   \\
\d E_{\vphi 2}&=&b^0{}_\vphi\d H_0+\d b^0 H_{0\vphi}=
  \big(b^0{}_\vphi\d H_{0\th\vphi}
       +\d b^0{}_\vphi H_{0\th\vphi}\big)d\th d\vphi               \nn\\
  &=&\d\big(b^0{}_\vphi H_{0\th\vphi}\big)d\th d\vphi
          =4a_1\d\Big(\frac{ma}{\a^2}\Big)d\Om'\,.
\eea
\esubeq
Summing up the two terms and using $A_0=a_1$, one obtains
\be
E_\vphi=16\pi A_0\d\Big(\frac{ma}{\a^2}\Big)=E_\phi\,.          \lab{4.6}
\ee
The last equality follows from the trivial coordinate transformation $\xi_\phi=\xi_\vphi$, see \eq{4.2}.

\subsection{Energy}

Going over to the energy, we represent the expression
$\d E_t:=\d\G_\infty(\pd_t)$ by the sum of
\bea
&&\d E_{t1}=\frac{1}{2}\om^{ij}{}_t\d H_{ij}
                        +\frac{1}{2}\d\om^{ij}H_{ijt}\, ,            \nn\\
&&\d E_{t2}=b^i{}_t\d H_{i}+\d b^iH_{it}\,.
\eea
The nonvanishing contributions to $\d E_{t1}$ are
\bsubeq
\bea
&&\d\om^{12} H_{12t}=(\d\om^{12}{}_\th H_{12t\vphi})d\th d\vphi
                    =-A_0m\frac{\d f}{\a f}\sin\th d\th d\vphi\, ,   \nn\\
&&\d\om^{13}H_{13t}=(-\d\om^{13}{}_\vphi H_{13t\th})d\th\d\vphi
             =-A_0m\frac{2f\d\a-\a\d f}{\a^2 f}\sin\th d\th d\vphi\,,\nn\\
\Ra&&\d E_{t1}=-2A_0m\frac{\d\a}{\a^2}
                  =2A_0 m\d\Big(\frac{1}{\a}\Big)\times 4\pi\,.
\eea
In a similar manner,
\bea
&&b^0{}_t\d H_0=(b^0{_t}\d H_{0\th\vphi})d\th d\vphi
  =4a_1\frac{\a\d m-m\d\a}{\a^2}\sin\th d\th d\vphi\, ,              \nn\\
\Ra&&\d E_{t2}=4a_1\d\Big(\frac{m}{\a}\Big)\times 4\pi\, .
\eea
\esubeq
Thus, the complete result takes the form
\be
\d E_t=16\pi A_0\left[\frac{m}{2}\d\Big(\frac{1}{\a}\Big)
                +\d\Big(\frac{m}{\a}\Big)\right]\,,                 \lab{4.9}
\ee
which shows why Boyer-Lindquist coordintes are inadequate.  Namely, if \eq{4.9} were the final result, the variational equation for energy would not be integrable, and consequently, energy would not be even defined. As we noted earlier, the correct result can be obtained only by going over to the untwisted $(T,\phi)$ coordinates. Indeed, using the transformation law \eq{4.2}$_1$ for the components of $\xi$, the expression for $\d E_t=\d\G_\infty(\pd_t)$ is transformed into the final result for $\d E_T:=\d\G_\infty(\pd_T)$, given by
\be
\d E_T=\d E_t+\l a\d E_\vphi=16\pi A_0\d\Big(\frac{m}{\a^2}\Big)\,. \lab{4.10}
\ee

The results \eq{4.10} and \eq{4.6} for the asymptotic charges $E_T$ and $E_\phi$, respectively, coincide with those obtained by Hecht and Nester \cite{kads3}; in the GR limit, they reduce to the form found earlier by Henneaux and Teitelboim \cite{hentei}, see also Ref. \cite{bc3}.

\section{Entropy}\label{sec5}
\setcounter{equation}{0}

Entropy is defined  by the variational equation for $\G_H(\xi)$, with
\bea
&&\xi:=\pd_T-\Om_+\pd_\phi=\pd_t-\om_+\pd_\vphi\,,                  \nn\\
&&\om_+=\frac{a\a}{r_+^2+a^2}\,,\qquad
  \Om_+=\om_++\l a=\frac{a(1+\l r_+^2)}{r_+^2+a^2}\,.
\eea
In the analysis of $\d\G_H(\xi)$, the following relations are very useful:
\bea
&&N\pd_r N\big|_{r_+}=\frac{\kappa(r_+^2+a^2)}{\r_+^2}\, ,\qquad
  N\d N \big|_{r_+}=0\, ,                                           \nn\\
&&\xi\inn b^0\big|_{r_+}=N\frac{\r_+^2}{r_+^2+a^2}\,,
 \qquad\xi\inn b^a\big|_{r_+}=0\,.                                   \nn
\eea
They allow us to easily obtain the interior products $\xi\inn\a\equiv \a_\xi$ for any form $\a$ expressed in the orthonormal basis. Thus, for instance, using the expressions for the Riemannian connection $\tom^{ij}$ displayed in Appendix \ref{appA}, one finds
\bea
&&\xi\inn\tom^{01}=-N'(\xi\inn b^0)=-\k\, ,\qquad
      \xi\inn \tom^{02}=\frac{Na^2\sin\th\cos\th}{P(r_+^2+a^2)}\,,   \nn\\
&&\xi\inn\tom^{13}=-\frac{Nar_+}{P(r_+^2+a^2)}\sin\th\,,\qquad
\xi\inn\tom^{03}=\xi\inn\tom^{12}=0\,,\quad \xi\inn\tom^{23}\sim N^2\,.\nn
\eea
In a similar manner, one can calculate the interior products $\xi\inn\om^{ij}$, $\xi\inn H_{ij}$, and $\xi\inn H_i$, appearing in the variational equation \eq{2.4}.

In order to make our analysis of entropy as transparent as possible, we organize the calculations in several simpler steps.

\subsection{The basic result}

We begin with the calculation of the expression $\d\G_H(\xi)$, given in Eq. \eq{2.4}, by dividing it into two parts, denoted symbolically by $\d\G_1$ and
$\d\G_2$.

\subsubsection*{\mb{\d\G_1=\frac{1}{2}\om^{ij}{}_\xi\d H_{ij}
                         +\frac{1}{2}\d\om^{ij}H_{ij\xi}}}

The only nonvanishing contributions stemming from the first element of $\d\G_1$ are
\bsubeq\lab{5.2}
\bea
&&\om^{01}{}_\xi\d H_{01} ~[=]~ \om^{01}{}_\xi\d H_{01\th\vphi}
  =2\bA_0\Big(\k-V_1\frac{\r_+^2}{r_+^2+a^2}\Big)
              \d\Big(\frac{r_+^2+a^2}{\a}\Big)\sin\th\,,          \lab{5.2a}\\
&&\om^{03}{}_\xi\d H_{03}+\om^{13}{}_\xi\d H_{13}
  ~[=]~ K^{03}{}_\xi\,\d(H_{03\th\vphi}+H_{13\th\vphi})
                                  +\tom^{13}{}_\xi\d H_{13\th\vphi}  \nn\\
&&\hspace{20pt}
  =2\bA_0\Big(\frac{1}{N}V_3\frac{\r_+^2}{r_+^2+a^2}\Big)\cdot
    \d\Big(PN\frac{a}{\a}\Big)\sin^2\th
   +2\l b_4\frac{ar_+N}{P(r_+^2+a^2)}
     \d\Big(\frac{mr_+}{N\r_+^2}\frac{Pa}{\a}\Big)\sin^3\th\,.       \nn\\
                                                                  \lab{5.2b}
\eea
\esubeq
Here, the symbol $[=]$ stands for an equality up to the factor $d\th d\vphi$, and $\bA_0=a_0-\l b_6$. In $\d H_{13\th\vphi}$, the term proportional to $\bA_0$ is omitted as it vanishes on horizon, $N\d N|_{r_+}=0$.

In the second element of $\d\G_1$ there are $2+2$ nonvanishing contributions,
\bsubeq\lab{5.3}
\bea
&&\d\om^{02}H_{02\xi}+\d\om^{12}H_{12\xi} ~[=]~
   \d\tom^{12}{}_\th H_{12\xi\vphi}
  +\d K^{02}{}_\th(H_{02\xi\vphi}+H_{12\xi\vphi})                    \nn\\
&&\hspace{20pt}
   =-2\bA_0\d\Big(\frac{mPr_+^2}{N\r_+^4}\Big)\frac{N\r_+^2}{P\a}\sin\th
   -2\l b_4\d\Big(\frac{NPr_+}{\r_+^2}\Big)\frac{mr_+}{NP\a}\sin\th\,,
                                                                  \lab{5.3a}
\eea
and
\bea
&&\d\om^{03}H_{03\xi}+\d\om^{13}H_{13\xi}~[=]~
  \approx-\d K^{03}{}_\vphi (H_{03\xi\th}+H_{13\xi\th})
                       -\d\tom^{13}{}_\vphi H_{13\xi\th}              \nn\\
&&\hspace{20pt}
  =-2\bA_0\d\Big(\frac{mr_+^2}{NP\r_+^2\a}\Big)
               \frac{NP\r_+^2}{r_+^2+a^2}\sin\th
   -2\l b_4\d\Big(\frac{Nr_+}{\a P}\Big)
           \frac{mr_+}{N}\frac{P}{r_+^2+a^2}\sin\th\,.            \lab{5.3b}
\eea
\esubeq
In $H_{13\xi\th}$, the term proportional to $\bA_0$ is omitted.

\subsubsection*{\mb{\d\G_2=b^i{}_\xi\d H_i+\d b^iH_{i\xi}}}

The only nonvanishing contributions from $\d\G_2$ are
\bsubeq\lab{5.4}
\bea
&&b^0{}_\xi\d H_0~[=]~ b^0{}_\xi \d H_{0\th\vphi}=N\frac{\r_+^2}{r_+^2+a^2}
  \d\Big[\frac{2a_1mr_+^2}{N\a\r_+^4}(r_+^2+a^2+\r_+^2)\Big]
                                                      \sin\th\,,  \lab{5.4a}\\
&&\d b^0 H_{0\xi}~[=]~-\d b^0{}_\vphi H_{0\xi\th}
  =-2a_1\d\Big(\frac{Na}{\a}\Big)
      \frac{V_3P}{N}\frac{\r_+^2}{r_+^2+a^2}\sin^2\th\,,         \lab{5.4b}\\
&&\d b^2 H_{2\xi}~[=]~\d b^2{}_\th H_{2\xi\vphi}-\d b^2{}_\vphi H_{2\xi\th}
  =2a_1(\d P)(V_1-V_5)\frac{\sin\th}{P\a}\r_+^2\,,               \lab{5.4c}\\
&&\d b^3 H_{3\xi}~[=]~-\d b^3{}_\vphi H_{3\xi\th}
  =2a_1\d\Big(\frac{r_+^2+a^2}{P\a}\Big)
            (V_1-V_5)P\frac{\r_+^2}{r_+^2+a^2}\sin\th\,.         \lab{5.4d}
\eea
\esubeq

\subsection{Simplifications}\label{sub52}

The expressions for entropy found in \eq{5.2}--\eq{5.4} look rather complex. It is almost evident that prior to any direct calculation,  they should be simplified. The evidence for the existence of the following two simplifications is provided in Appendix \ref{appB}:
\bitem
\item[\mb{T1.}] The sum of the terms proportional to $\d N/N$ in \eq{5.2}-\eq{5.4} vanishes.\vsm
\item[\mb{T2.}] The sum of the terms proportional to $\d P/P$ in \eq{5.2}-\eq{5.4} vanishes.
\eitem
As a consequence, the original expressions become notably simpler:
\bsubeq
\bea
\text{\eq{5.2a}}:&&
  2\bA_0\left[\k-V_1\frac{\r_+^2}{r_+^2+a^2}\right]\cdot
              \d\Big(\frac{r_+^2+a^2}{\a}\Big)\sin\th\,,            \nn\\
\text{\eq{5.2b}}:&&
    2\bA_0\Big(V_3P\frac{\r_+^2}{r_+^2+a^2}\Big)\cdot
                            \d\Big(\frac{a}{\a}\Big)\sin^2\th
    +2\l b_4\frac{ar_+}{(r_+^2+a^2)}
            \d\Big(\frac{mr_+}{\r_+^2}\frac{a}{\a}\Big)\sin^3\th\,.\qquad
                                                                 \lab{5.5a}
\eea
\bea
\text{\eq{5.3a}}:&&
  -2\bA_0\d\Big(\frac{mr_+^2}{\r_+^4}\Big)\frac{\r_+^2}{\a}\sin\th
  -2\l b_4\d\Big(\frac{r_+}{\r_+^2}\Big)\frac{mr_+}{\a} \sin\th\,,   \nn\\
\text{\eq{5.3b}}:&&
   -2\bA_0\d\Big(\frac{mr_+^2}{\r_+^2\a}\Big)
               \frac{\r_+^2}{r_+^2+a^2}\sin\th
   -2\l b_4\d\Big(\frac{r_+}{\a}\Big)
               \frac{mr_+}{r_+^2+a^2}\sin\th\,.                   \lab{5.5b}
\eea
\bea
\text{\eq{5.4a}}:&&
   2a_1\frac{\r_+^2}{r_+^2+a^2}
  \d\Big[\frac{mr_+^2}{\a\r_+^4}(r_+^2+a^2+\r_+^2)\Big]\sin\th\,,    \nn\\
\text{\eq{5.4b}}:&&
  -2a_1\d\Big(\frac{a}{\a}\Big)
                   V_3P\frac{\r_+^2}{r_+^2+a^2}\sin^2\th\,,          \nn\\
\text{\eq{5.4c}}:&& =0\,,                                            \nn\\
\text{\eq{5.4d}}:&&
   2a_1\d\Big(\frac{r_+^2+a^2}{\a}\Big)
        (V_1-V_5)\frac{\r_+^2}{r_+^2+a^2}\sin\th\,.               \lab{5.5c}
\eea
\esubeq
In further analysis, we shall use the relation  $\bA_0=A_0+\l b_4$ to express these results in terms of only \emph{two independent coupling constants}, $A_0$ and $\l b_4$. In this process, one should use the identity $a_1\equiv A_0$.

\subsection{The terms proportional to \mb{\l b_4}}

Since the contributions in \eq{5.5c} are proportional to $a_1\equiv A_0$, the $\l b_4$ contributions are determined by replacing $\bA_0\to\l b_4$ into \eq{5.5a} and \eq{5.5b}). Then, by dividing each term by $2\l b_4$ (for simplicity), one obtains
\bea
\text{\eq{5.2a}}:&&
   \left[\k-\frac{m(r_+^2-a^2\cos^2\th)}{\r_+^2(r_+^2+a^2)}\right]
                     \d\Big(\frac{r_+^2+a^2}{\a}\Big)\sin\th\,,      \nn\\
\text{\eq{5.2b}}:&&
   \frac{amr_+^2\sin^3\th}{\r_+^2(r_+^2+a^2)}
                      \d\Big(\frac{a}{\a}\Big)
   +\frac{ar_+}{r_+^2+a^2}
    \d\Big(\frac{mr_+}{\r_+^2}\frac{a}{\a}\Big)\sin^3\th\,,           \nn\\
\text{\eq{5.3a}}:&&
   -\Big[\frac{\r_+^2}{\a}\d\left(\frac{mr_+^2}{\r_+^4}\right)
   +\frac{mr_+}{\a}\d\Big(\frac{r_+}{\r_+^2}\Big)\Big]\sin\th\,,      \nn\\
\text{\eq{5.3b}}:&&
   -\Big[\frac{\r_+^2}{r_+^2+a^2}\d\Big(\frac{mr_+^2}{\a\r_+^2}\Big)
   +\frac{mr_+}{r_+^2+a^2}\d\Big(\frac{r_+}{\a}\Big)\Big]\sin\th\,. \lab{5.6}
\eea
These contributions can be further simplified, as shown in Appendix \ref{appB}.
\bitem
\item[\mb{T3.}]  When the sum of the terms in \eq{5.6} is integrated over $d\th d\vphi$, it vanishes.
\eitem
This result allows us to go over to the final stage of the analysis of entropy.

\subsection{The terms proportional to $A_0$}

The remaining contributions proportional to $A_0$ are obtained by the substitution $\bA_0\to A_0$ into \eq{5.5a} and \eq{5.5b}. By a suitable rearrangement, the result can be expressed as
\bea
&&\pha{\hspace{-2cm}}
  \text{\eq{5.2a}}+\text{\eq{5.2b}}_1+\text{\eq{5.3a}}_1+\text{\eq{5.3b}}_1:\nn\\
&&\pha{\hspace{-1cm}}
  2A_0\sin\th\left[\left(\k-\frac{V_1\r_+^2}{r_+^2+a^2}\right)
                                     \d\left(\frac{r_+^2+a^2}\a\right)
  +\frac{amr_+^2\sin^2\th}{\r_+^2(r_+^2+a^2)}\d\left(\frac a\a\right)\right.\nn\\
&&\pha{\hspace{4cm}}
  \left.-\frac{\r_+^2}\a\d\left(\frac{mr_+^2}{\r_+^4}\right)
  -\frac{\r_+^2}{r_+^2+a^2}\d\left(\frac{mr_+^2}{\a\r_+^2}\right)\right]\,,\nn\\
&&\pha{\hspace{-2cm}}
  \text{\eq{5.4a}}+\text{\eq{5.4b}}+\text{\eq{5.4c}}+\text{\eq{5.4d}}: \nn\\
&&\pha{\hspace{-1cm}}
  2a_1\frac{\r_+^2}{r_+^2+a^2}\sin\th\left[V_1\d\left(\frac{r_+^2+a^2}\a\right)
       -\frac{amr_+^2\sin^2\th}{\r_+^4}\d\left(\frac a\a\right)\right. \nn\\
&&\pha{\hspace{3.5cm}}
  \left.+\frac{r_+^2+a^2}\a\d\left(\frac{mr_+^2}{\r_+^4}\right)
   +\d\left(\frac{mr_+^2}{\a\r_+^2}\right)\right]\,.                   \nn
\eea
After using $A_0=a_1$, all these contributions sum up to a simple expression
\be
\text{\eq{5.2}}+\text{\eq{5.3}}+\text{\eq{5.4}}=
2A_0\k \sin\th\d\left(\frac{r_+^2+a^2}\a\right)\,.                   \lab{5.7}
\ee
Then, the integration over $d\th d\vphi$ yields the final result
\be
\d\G_H=8\pi A_0 \k\d\Big(\frac{r_+^2+a^2}{\a}\Big)=T\d S\, ,
       \qquad S:=16\pi A_0 \frac{\pi(r_+^2+a^2)}{4}\,,
\ee
where $T=\k/2\pi$ is the black hole temperature and $S$ the Kerr-AdS entropy in PG.

\section{The first law}\label{sec6}
\setcounter{equation}{0}

In the Hamiltonian approach described in section \ref{sec2}, the asymptotic charges and entropy are defined by the variational equations \eq{2.4} as a priory independent quantities. The results that we found for $\d E_T,\d E_\vphi$ and $\d\G_H$, combined with the identity derived in Appendix \ref{appC}, imply the validity of the first law of black hole thermodynamics for the Kerr-AdS black hole,
\be
T\d S=\d E_T-\Om_+\d E_\vphi\, ,                                   \lab{6.1}
\ee
in accordance with Eq. \eq{2.5}.

\section{Concluding remarks}\label{sec7}
\setcounter{equation}{0}

In the present paper, we performed a classical Hamiltonian analysis of  the thermodynamic variables, energy, angular momentum and entropy, for the Kerr-AdS spacetimes in PG.

Our analysis relies on the Kerr-AdS solution with torsion, constructed some thirty years ago by Baekler et al. \cite{kads1,kads2}. The results for energy and angular momentum coincide with those obtained by Hecht and Nester \cite{kads3}. In both their and our analyses, it was essential to understand the limitations of the Boyer-Lindquist coordinates at large distances in accordance with the ideas of Henneaux and Teitelboim \cite{hentei}, the ideas which can be traced back to the work of Carter \cite{carter}.

As far as we know, the result \eq{6.1} for entropy is completely new in the literature, although our earlier results for the spherically symmetric and asymptotically flat Kerr solutions \cite{bc1,bc2,bc3} led to certain ideas on what might be the answer in the Kerr-KAdS case. The calculations producing the final result for the Kerr-AdS entropy are rather complex, but at the end, they confirm that black hole entropy in PG can be interpreted as the canonical charge on horizon.

In spite of a very different geometric/dynamical content of PG and GR, our analysis shows that the related Kerr-AdS thermodynamic variables differ solely by a constant multiplicative factor. This somewhat puzzling situation may indicate the need for a deeper understanding of the role of boundary conditions.

\section*{Acknowledgments}

This work was partially supported by the Serbian Science Foundation under Grant No. 171031.

\appendix
\section{Technical aspects of Kerr-AdS solution}\label{appA}
\setcounter{equation}{0}

In this appendix, we present some detailed technical characteristics of the Kerr-AdS solution.

First, we display here the explicit form of the Riemannian Kerr-AdS connection \eq{3.7},
\bea
&&\tom^{01}=-N'b^0-\frac{ar}{P\r^2}\sin\th b^3\,,                    \nn\\
&&  \tom^{02}=\frac{a^2\sin\th\cos\th}{P\r^2}b^0
            -\frac{aN}{\r^2}\cos\th b^3\,,                           \nn\\
&&\tom^{03}=-\frac{ar}{P\r^2}\sin\th b^1
                 +\frac{aN}{\r^2}\cos\th b^2\,,                      \nn\\
&&\tom^{12}=\frac{a^2\sin\th\cos\th}{\r^2P}b^1
                                     +\frac{rN}{\r^2}b^2\,,          \nn\\
&&\tom^{13}=-\frac{ar}{P\r^2}\sin\th b^0+\frac{Nr}{\r^2}b^3\,,       \nn\\
&&\tom^{23}=-\frac{aN}{\r^2}\cos\th b^0
            +\frac{P\cos\th-\pd_\th P\sin\th}{P^2\sin\th}b^3\,.
\eea
Then, the irreducible components of the torsion 2-form \eq{3.8} are found to be
\bea
&&\ir{2}T^0=\ir{2}T^1=\frac{1}{3N}(-V_1+2V_5)b^0b^1\, ,              \nn\\
&&\ir{2}T^c=\frac{1}{3N}\big(-V_1+2V_5\big)b^-b^c\, ,
                                           \qquad c=(2,3),           \nn\\
&&\ir{1}T^0=\ir{1}T^1=
  -\frac{1}{N}\Big[\frac{2}{3}(V_1+V_5)b^0b^1+2V_4 b^2b^3\Big]
  +\frac{1}{N^2}b^-V_cb^c\, ,                                     \nn\\
&&\ir{1}T^2=\frac{1}{N}\Big[
            \frac{1}{3}(V_1+V_5)b^-b^2+V_4b^-b^3\Big]\,,   \nn\\
&&\ir{1}T^3=\frac{1}{N}\Big[
            \frac{1}{3}(V_1+V_5)b^-b^3-V_4b^-b^2\Big]\,.   \nn\\
&&\ir{3}T^i=0\,.
\eea
Finally, the explicit forms of the covariant momenta read
\bea
&&H_{01}=-2\bA_0 b^2b^3\, ,                                     \nn\\
&&H_{02}=2\bA_0 b^1b^3+2b_4\frac{\l m r}{\D}b^-b^3\,,        \nn\\
&&H_{12}=-2\bA_0 b^0b^3-2b_4\frac{\l m r}{\D}b^-b^3\,,       \nn\\
&&H_{03}=-2\bA_0 b^1b^2-2b_4\frac{\l mr}{\D}b^-b^2\,,        \nn\\
&&H_{13}=2\bA_0 b^0b^2+2b_4\frac{\l m r}{\D}b^-b^2\,,        \nn\\
&&H_{23}=-2\bA_0 b^0b^1\, ,
\eea
\bea
&&H_0=-H_1=\frac{4a_1}{N}\Big[-V_4b^0b^1+V_5b^2b^3\Big]
      +\frac{2a_1}{N^2}\Big[b^-(-V_2b^3+V_3b^2)\Big]\,,\nn\\
&&H_2=-\frac{2a_1}{N}\Big[(-V_1+V_5)b^-b^3+V_4b^-b^2\Big]\,,\nn\\
&&H_3=-\frac{2a_1}{N}\Big[(V_1-V_5)b^-b^2+V_4b^-b^3\Big]\,,
\eea

\section{On the evaluation of entropy}\label{appB}
\setcounter{equation}{0}

In this appendix, we discuss certain technical details of the derivation of entropy given in the main text.

\subsection{Elimination of $\d N/N$ and \mb{\d P/P} terms}

Starting from the basic results on entropy obtained in Eqs. \eq{5.2}-\eq{5.4}, we are now going to show that both $\d N/N$ and $\d P/P$ terms cancel out.

Consider first the coefficients of the $\d N/N$ terms. By a suitable rearrangement of these coefficients, shown in the following formulas
\bsubeq
\bea
\text{\eq{5.3a}}_1+\text{\eq{5.3b}}_1:&&
  2\bA_0\frac{mr_+^2}{\a\r_+^2}\Big(1+\frac{\r_+^2}{r_+^2+a^2}\Big)
                                                        \sin\th\,,  \nn\\
\text{\eq{5.3a}}_2+\text{\eq{5.3b}}_2:&&
 -2\l b_4\frac{mr_+^2}{\a\r_+^2}\Big(1+\frac{\r_+^2}{r_+^2+a^2}\Big)
                                                        \sin\th\,,  \nn\\
\text{\eq{5.4a}}:&&
  -2a_1\frac{mr_+^2}{\a\r_+^2}\Big(1+\frac{\r_+^2}{r_+^2+a^2}\Big)
                                                        \sin\th\,, \nn
\eea
one can directly conclude that their sum vanishes, as a consequence of $\bA_0\equiv a_1+\l b_4$. There are two more contributions of this type,
\bea
\text{\eq{5.2b}}:&&
   2(\bA_0-\l b_4)\frac{m r_+^2a^2}{\a\r_+^2(r_+^2+a^2)}\sin^3\th\,, \nn\\
\text{\eq{5.4b}}:&&
  -2a_1\frac{m r_+^2a^2}{\a\r_+^2(r_+^2+a^2)}\sin^3\th\,,             \nn
\eea
\esubeq
whose sum also vanishes. Hence, all $(\d N)/N$ terms in entropy can be simply ignored.

A similar analysis shows that the sum of all $\d P/P$ terms also vanishes:
\bea
\text{\eq{5.2b}}_1+\text{\eq{5.3a}}_1+\text{\eq{5.3b}}_1:&&
  2\bar A_0\frac{mr_+^2\sin\th}{\a}
     \left(\frac{a^2\sin^2\th}{\rho_+^2(r_+^2+a^2)}
          -\frac1{\r_+^2}+\frac1{r_+^2+a^2}\right)=0\,,             \nn\\
\text{\eq{5.2b}}_2+\text{\eq{5.3a}}_2+\text{\eq{5.3b}}_2:&&
  2\l b_4\frac{mr_+^2\sin\th}{\a}
     \left(\frac{a^2\sin^2\th}{\rho_+^2(r_+^2+a^2)}
          -\frac1{\r_+^2}+\frac1{r_+^2+a^2}\right)=0\,,             \nn\\
\text{\eq{5.4c}}+\text{\eq{5.4d}}:&&
     2a_1(V_1-V_5)\frac{\sin\th}{\a}\r_+^2                          \nn\\
   &&\qquad -2a_1\frac{r_+^2+a^2}{\a}
               (V_1-V_5)\frac{\r_+^2}{r_+^2+a^2}\sin\th=0\,.         \nn
\eea

\subsection{Elimination of \mb{\l b_4} terms}

Let us now analyze Eq. \eq{5.6} from the main text, which is focused on the contributions from the $\l b_4$ terms. In order to simplify the formulas, we \emph{temporarily} omit the common factor $\sin\th$ and rewrite the result in a more convenient form:
\bea
\text{\eq{5.2a}}:&&
    \left[\k-\frac{m(r_+^2-a^2\cos^2\th)}{\r_+^2(r_+^2+a^2)}\right]
                        \d\Big(\frac{r_+^2+a^2}{\a}\Big)\,,      \nn\\
\text{\eq{5.2b}}:&& \frac{ar_+\sin^2\th}{r_+^2+a^2}\left[
        2\frac{mr_+}{\r_+^2}\d\Big(\frac{a}{\a}\Big)
      +\frac{a}{\a}\d\Big(\frac{mr_+}{\r_+^2}\Big)\right]\,,     \nn\\
\text{\eq{5.3a}}:&&-\frac{\r_+^2}{\a}\left[
     \frac{r_+}{\r_+^2}\d\Big(\frac{mr_+}{\r_+^2}\Big)
     +2\frac{mr_+}{\r_+^2}\d\Big(\frac{r_+}{\r_+^2}\Big)\right]\,,\nn\\
\text{\eq{5.3b}}:&&-\frac{\r_+^2}{r_+^2+a^2}
     \left[\frac{r_+}{\a}\d\Big(\frac{mr_+}{\r_+^2}\Big)
     +2\frac{mr_+}{\r_+^2}\d\Big(\frac{r_+}{\a}\Big)\right]\,.\lab{B.2}
\eea
Now, if the first term in \eq{5.2a} is replaced by
\be
\k\d A=\frac{2r_+^2}{\a(r_+^2+a^2)}\d m
              +\frac{2am(-1+3\l r_+^2)}{\a^2 (r_+^2+a^2)}\d a\,,
\ee
one can directly conclude that
\bitem
\item the sum of all $\d m$ terms in \eq{B.2} vanishes.
\eitem
As a consequence, one can further simplify the form of \eq{B.2}.
By rearranging the last 3 lines, \eq{B.2} becomes
\bea
\text{\eq{5.2a}}:&&
  \frac{2am(-1+3\l r_+^2)}{\a^2 (r_+^2+a^2)}\d a
  -\frac{m(r_+^2-a^2\cos^2\th)}{\r_+^2(r_+^2+a^2)}
                       \d\Big(\frac{r_+^2+a^2}{\a}\Big)\,,       \nn\\
\text{\eq{5.2b}}_2+\text{\eq{5.3a}}+\text{\eq{5.3b}}_1:&&
\frac{mr_+}{\a(r_+^2+a^2)}\d\Big(\frac{r_+}{\r_+^2}\Big)
    \Big[-2 \r_+^2-2(r_+^2+a^2)\Big]\,,                          \nn\\
\text{\eq{5.2b}}_1+\text{\eq{5.3b}}_2:&&
  2\frac{mr_+}{r_+^2+a^2}\left[
  \frac{ar_+\sin^2\th}{\r_+^2}\d\Big(\frac{a}{\a}\Big)
  -\d\Big(\frac{r_+}{\a}\Big)\right]\, .                       \lab{B.4}
\eea

\subsubsection*{The \mb{\d r_+} terms in \eq{B.4} vanish}

In Eqs. \eq{B.4}, one can treat $\d r_+$ and $\d a$ as two independent variations on horizon. Consider first the $\d r_+$ part of \eq{B.4}, defined by $\d a=0$, but with $a\ne 0$. Then, by integrating \eq{B.4}$\times\d\Om$, where $d\Om\equiv\sin\th d\th d\vphi$, one finds that the sum of these terms vanishes.

\subsubsection*{The \mb{\d a} terms in \eq{B.4} vanish}

The remaining, explicit $\d a$ terms in \eq{B.4} are given by
\bsubeq\lab{B.5}
\bea
&& \frac{2am(-1+3\l r_+^2)}{\a^2 (r_+^2+a^2)}\d a
  -\frac{m(r_+^2-a^2\cos^2\th)}{\r_+^2(r_+^2+a^2)}
                       \rd\Big(\frac{r_+^2+a^2}{\a}\Big) \,,          \nn\\
&&-\frac{mr_+}{\a(r_+^2+a^2)}\rd\Big(\frac{r_+}{\r_+^2}\Big)
    \Big[2 \r_+^2+2(r_+^2+a^2)\Big]\,,                                \nn\\
&&2\frac{mr_+}{r_+^2+a^2}\left[
  \frac{ar_+\sin^2\th}{\r_+^2}\rd\Big(\frac{a}{\a}\Big)
  -\rd\Big(\frac{r_+}{\a}\Big)\right]\, ,                         \lab{B.5a}
\eea
where the variation $\rd X$ acts only on $a$'s that are
\emph{explicitly} present in $X$,
\bea
&&\rd\Big(\frac{r_+^2+a^2}{\a}\Big) =\frac{2a(1+\l r^2)}{\a^2}\d a\,,\nn\\
&&\rd\Big(\frac{r_+}{\r_+^2}\Big)
  =-\frac{r_+}{\r_+^4}\big(2a\cos^2\th\big)\d a\, ,                  \nn\\
&&\rd\Big(\frac{a}{\a}\Big)=\frac{1+\l a^2}{\a^2}\d a\,,\qquad
  \rd\Big(\frac{r_+}{\a}\Big)=\frac{2\l ar_+}{\a^2}\d a\,.       \lab{B.5b}
\eea
\esubeq
A direct integration of the terms in \eq{B.5a}$\times d\Om$ shows that their sum vanishes.

To summarize:
\bitem
\item The sum of the $\l b_4$ terms in \eq{B.4}$\times d\Om$ vanishes after integration.
\eitem
\section{Elementary 1st law as an identity}\label{appC}
\setcounter{equation}{0}

Here, we consider an ``elementary" version of the first law. Lets us define
\be
\cS:=\frac{r_+^2+a^2}{\a}\,,\quad M:=\frac{m}{\a^2}\,,\qquad J:=Ma\, .
\ee
By calculating $\d\cS$ as a function of $\d r_+$ and $\d a$, one can use the horizon equation to express $\d r_+$ in terms of $\d m$ and $\d a$, which yields
\bea
&&\k\d r_+=\frac{r_+}{r_+^2+a^2}\d m
            -\frac{a(1+\l r_+^2)}{r_+^2+a^2}\d a\, ,                 \nn\\
&&\frac{\k}{2}\d\cS=\frac{r_+^2}{\a(r_+^2+a^2)}\d m
                 +\frac{a(1+\l r_+^2)(-1+3\l r_+^2)}{2\a^2 r_+}\d a\, .  \lab{A.2}
\eea
Then, after calculating the variation of the charge on horizon,
\be
\d M-\Om\d J=\frac{r_+^2}{\a(r_+^2+a^2)}\d m
                +\frac{am(-1+3\l r_+^2)}{\a^2(r_+^2+a^2)}\d a\,,
\ee
one obtains the relation
\be
\frac{\k}{2}\d\cS = \d M-\Om\d J~~\text{~~ on horizon}\,.           \lab{C.4}
\ee
This identity is an elementary version of the first law, determined solely from the definition of horizon.


\end{document}